\documentclass[fleqn,12pt,twoside]{article}
\usepackage[headings]{espcrc1}

\readRCS
$Id: espcrc1.tex,v 1.2 2004/02/24 11:22:11 spepping Exp $
\usepackage{graphicx}

\newcommand{\AmS}{{\protect\the\textfont2
  A\kern-.1667em\lower.5ex\hbox{M}\kern-.125emS}}

\newcommand{\gev}{GeV/$c$}
\newcommand{\pt}{p_{T}}

\newcommand{\dphi}{\Delta\phi}
\newcommand{\deta}{\Delta\eta}

\title{Hard Probes are the ``Meridian Line'', sQGP is the Forbidden City}

\author{Fuqiang Wang\address[MCSD]{Department of Physics, Purdue University,
    525 Northwestern Avenue, Indiana, USA}
       }

\runtitle{Hard Probes are the ``Meridian Line'', sQGP is the Forbidden City}
\runauthor{F. Wang}

\begin{document}
\maketitle

\begin{abstract}
Relativistic heavy-ion collisions at BNL/RHIC have created a hot and dense nuclear matter, the strongly interacting Quark-Gluon Plasma (sQGP). Hard probes, high transverse momentum particles and jets, interact with the sQGP medium and lose energy. The interactions modify the properties of the hard probes; the medium responds to the energy loss of those probes in a collective way. Many properties of the sQGP have been learned from the modifications to the probes and the medium responses. I will review jet-like correlation results from RHIC, especially those measured by three-particle correlations, and discuss some of the fascinating physics.
\end{abstract}

\section{Introduction}

First, a few words on the title. The Forbidden City, everyone is familiar with. It is the center of China, literally ``The Middle Kingdom'', in ancient Chinese dynasties. The ``Meridian Line'' is the central path of the Forbidden City; it is the center of the center of the center of the world, as they say. The Meridian Line path is the widest at the Meridian Gate, the entrance to the Forbidden City. Every morning, streams of tourists rushing into the Meridian Gate would shake the crowd already inside; they would emerge not in the form of the original streams, but as scattered individuals from all over the places. The Meridian path becomes narrower and ``weaker'' into the interior of the Forbidden City, leading to many small, spreading paths. The Meridian Line is hardly recognizable at the back side of the Forbidden City. 

Much of these resembles collisions at Brookhaven National Laboratory's Relativistic Heavy-Ion Collider (BNL/RHIC). Relativistic heavy-ion collisions create a hot and dense medium of nuclear matter, the strongly coupled Quark-Gluon Plasma (sQGP). High transverse momentum ($\pt$) particles and jets, collectively called hard probes, are self-generated by underlying parton-parton hard scatterings in these collisions. Those energetic and ``strong'' hard probes interact with the sQGP and were found to lose energy becoming weaker and weaker while traversing the sQGP. As a result, fewer high-$\pt$ particles were measured in central heavy-ion collisions than expected from elementary $pp$ collisions and the perturbative Quantum Chromodynamics theory~\cite{wp1,wp2}. Pictorially, the hard probes look just like the Meridian Line, and the created sQGP is the Forbidden City.

Hard-scattered partons emerge back-to-back on the transverse plane as collimated jets of hadrons. This was observed in $pp$ and d+Au collisions by dihadron azimuthal correlations with high-$\pt$ trigger particles~\cite{ISR,wp1,wp2} (see Fig.~\ref{fig:dphi}). In relativistic heavy-ion collisions two novel phenomena, not present in $pp$ or d+Au collisions, were observed -- the cone and the ridge. The cone and the ridge are two of the most fascinating findings at RHIC. I will review the experimental results on the cone and the ridge, focusing on three-particle correlations, and discuss what we have learned from them about the properties of the sQGP.

\section{The Cone}

\begin{figure}[htb]
\begin{minipage}{0.4\textwidth}
\includegraphics[width=\textwidth]{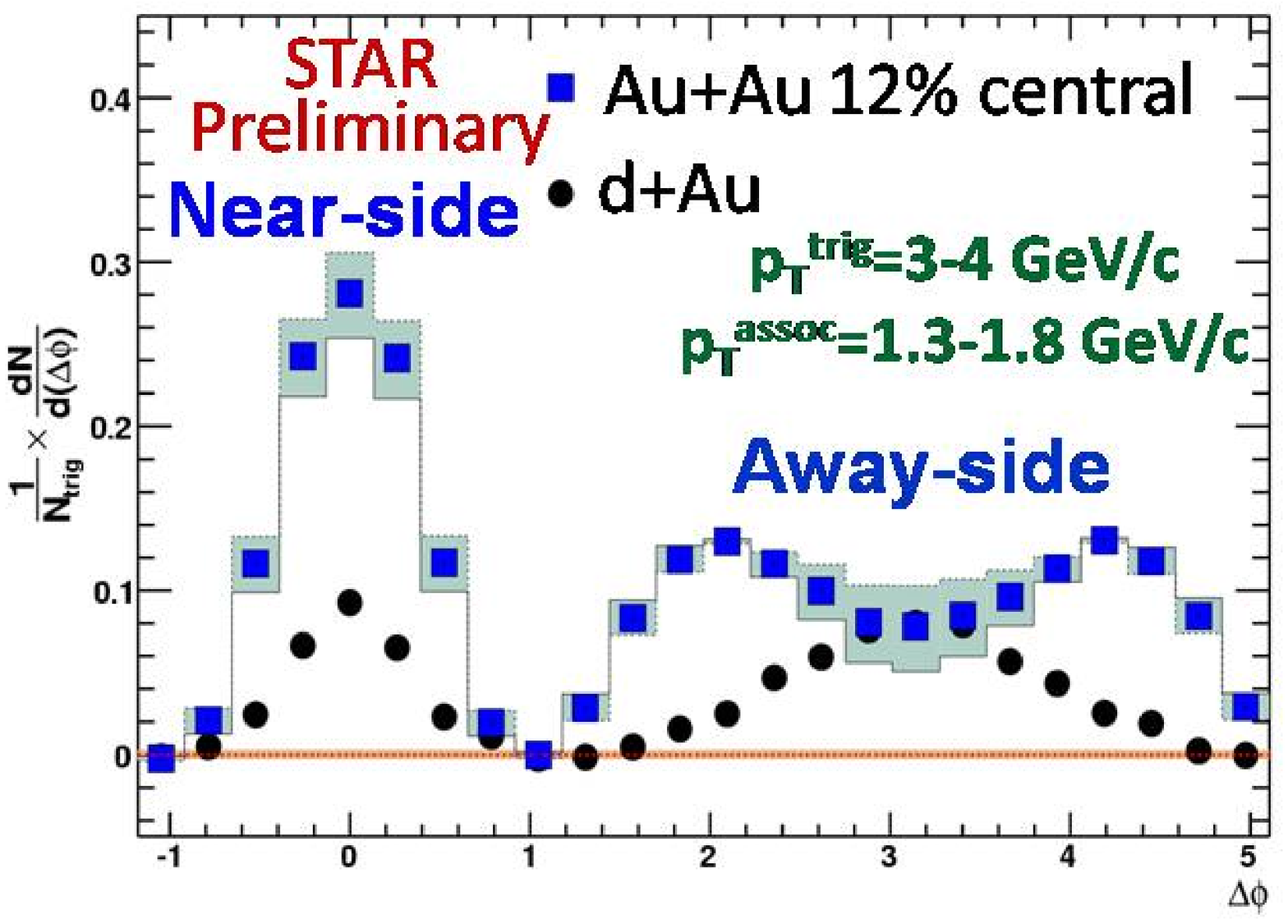}
\vspace*{-0.6in}
\caption{Dihadron correlation from STAR~\cite{Horner}. Flow background is normalized in $\dphi\approx 1$ by {\sc zyam} (zero yield at minimum) and subtracted.}
\label{fig:dphi}
\end{minipage}
\hspace*{0.02\textwidth}
\begin{minipage}{0.56\textwidth}
\includegraphics[width=0.555\textwidth]{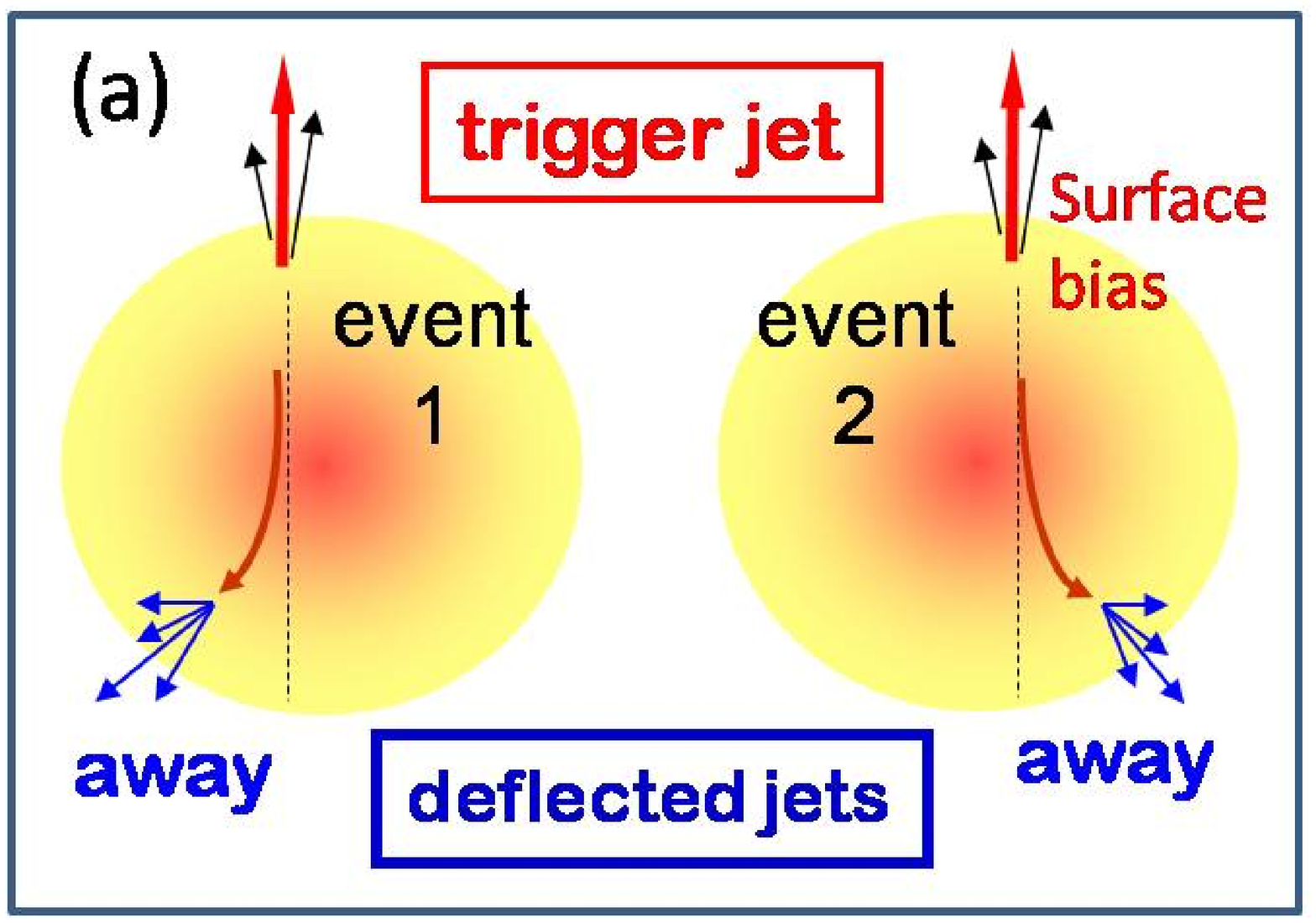}
\includegraphics[width=0.425\textwidth]{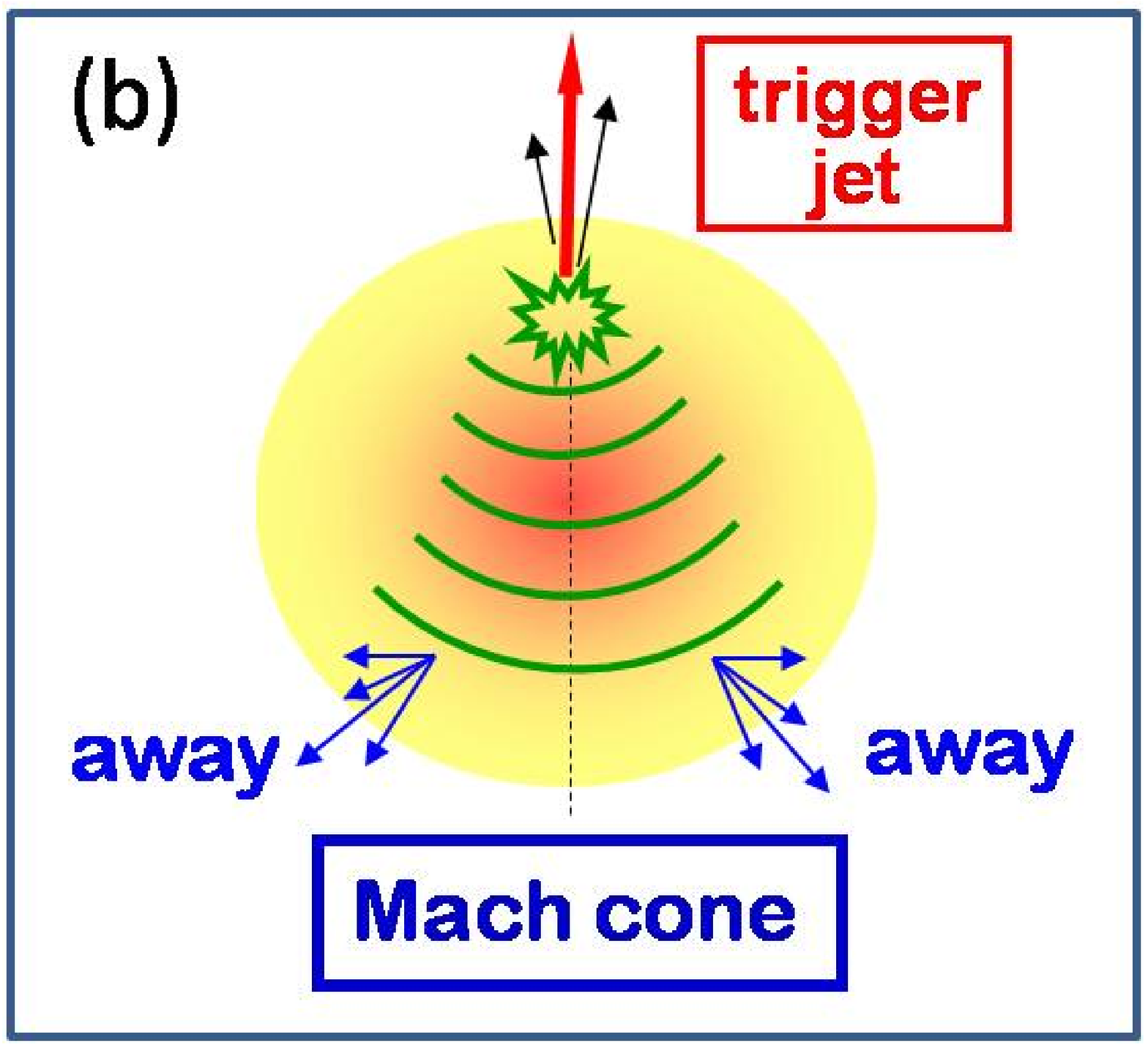}
\caption{Schematic diagrams of possible physics ecenarios for the away-side double-peak structure. (a) Medium deflected jets. (b) Mach-cone supersonic shock-waves.}
\label{fig:cartoon}
\end{minipage}
\end{figure}

Due to partonic energy loss, high-$\pt$ particles in relativistic heavy-ion collisions measured in detector are originated primarily from the surface of the collision zone. The away-side parton partner, opposite to a high-$\pt$ trigger particle, has to traverse the entire medium created in the collision. Dihadron correlations with high-$\pt$ trigger particles are found to be very broad on the away side, and even double-peaked for selected kinematic regions (see Fig.~\ref{fig:dphi}). Several physics mechanisms have been proposed, including deflected jets due to collective radial flow of the bulk~\cite{Armesto} or pathlength dependent energy loss~\cite{Hwa}, and conical emission due to $\check{\rm C}$erenkov gluon radiation~\cite{Dremin,Koch} or Mach-cone shock waves generated by energy deposition in the hydrodynamic medium~\cite{Greiner,Stoecker,Casalderrey,Ruppert,Renk}. Figure~\ref{fig:cartoon} depicts schematic diagrams of the deflected jets scenario and the Mach-cone scenario. Because experimental measurements are an average over many events, all these scenarios produce qualitatively the same dihadron correlation. In order to discriminate them, The STAR (Solenoidal Tracker at RHIC) experiment has carried out a three-particle azimuthal correlation analysis~\cite{3part}.

\begin{figure}[htb]
\begin{minipage}{0.63\textwidth}
\includegraphics[width=\textwidth]{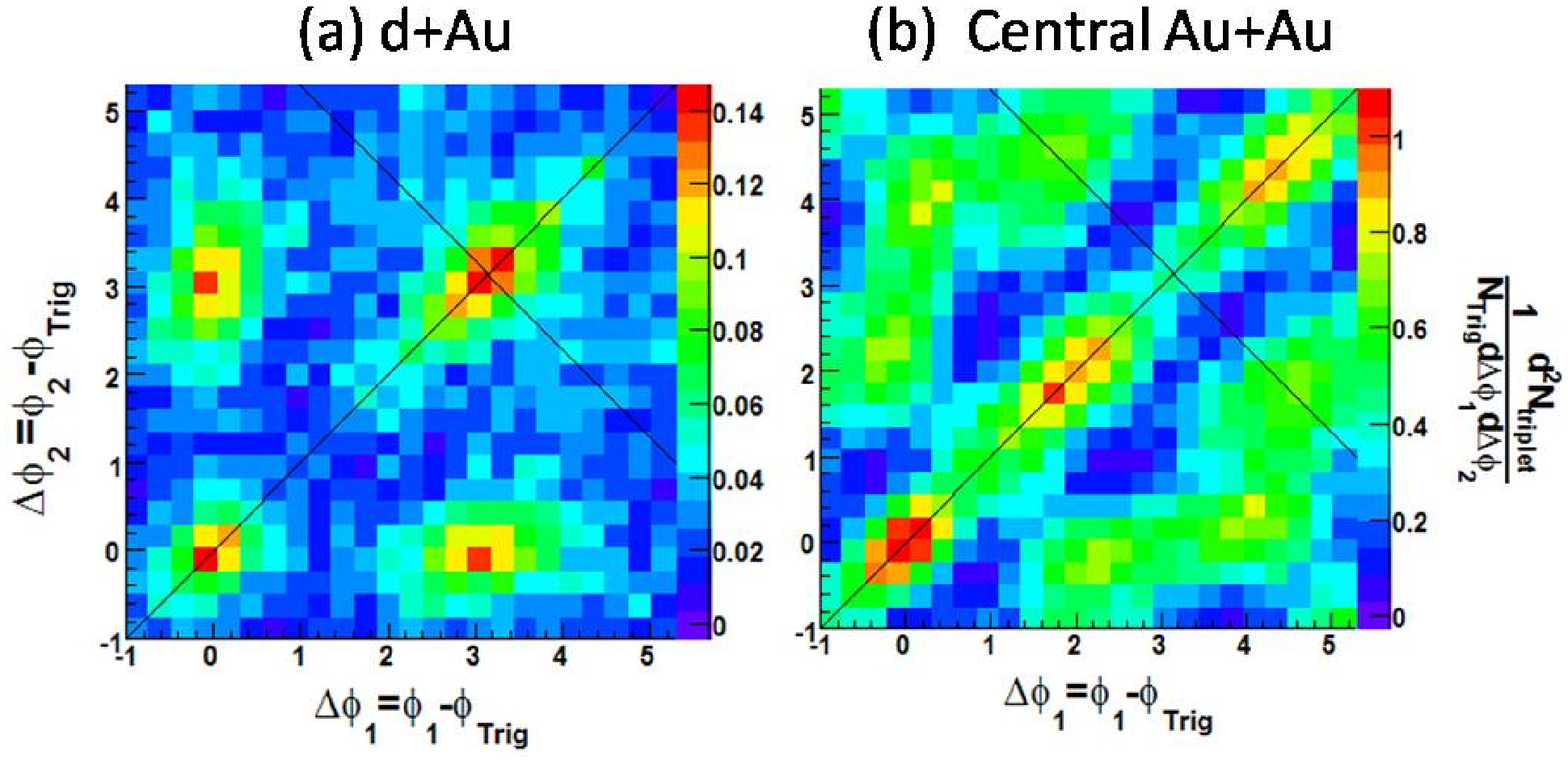}
\vspace*{-0.6in}
\caption{Three-particle azimuthal correlations in d+Au and central 12\% Au+Au collisions from STAR~\cite{3part}. Trigger and associated particle $\pt$ ranges are 3-4~\gev\ and 1-2~\gev, respectively.}
\label{fig:3part}
\end{minipage}
\hspace*{0.02\textwidth}
\begin{minipage}{0.35\textwidth}
\includegraphics[width=\textwidth]{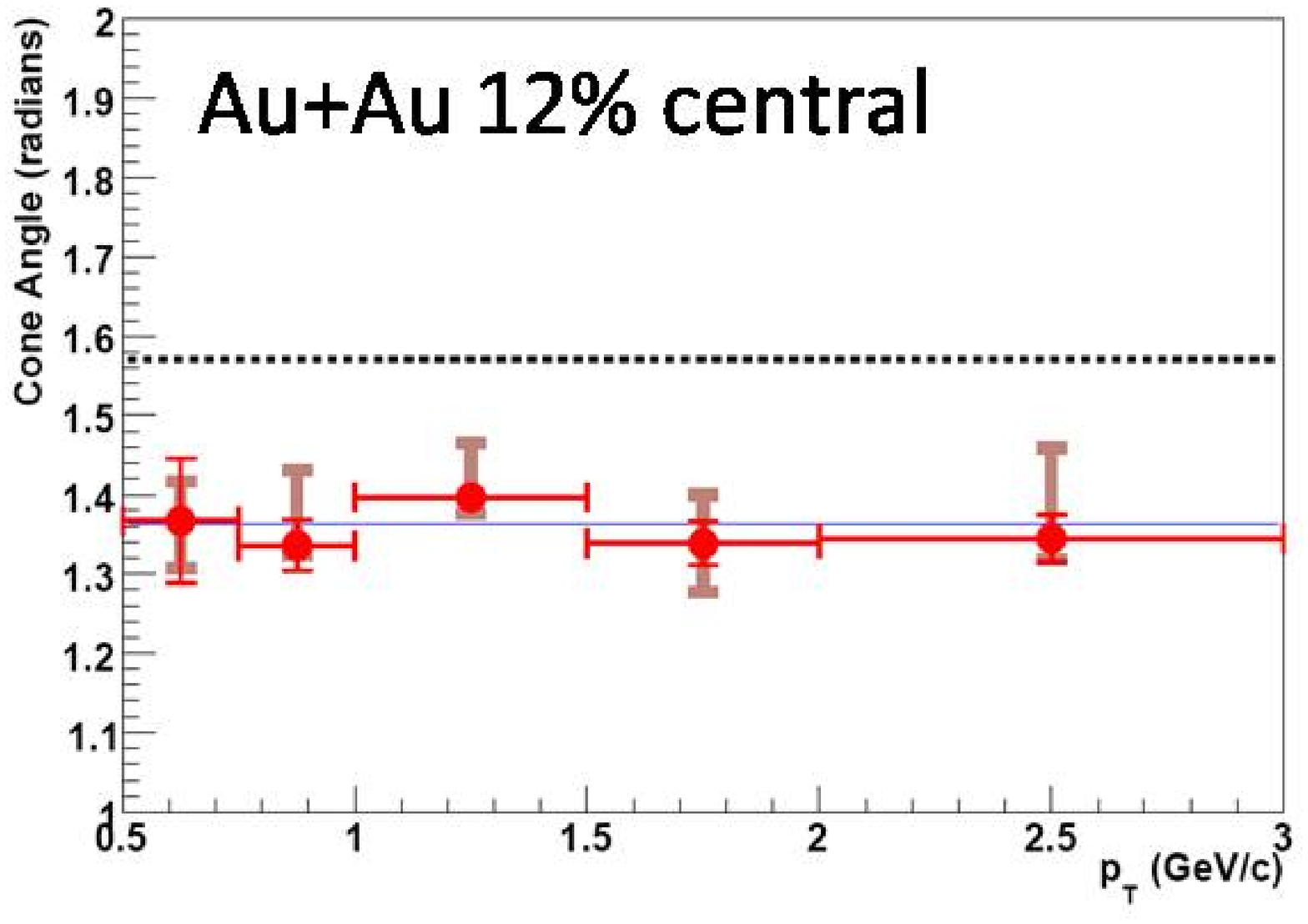}
\vspace*{-0.4in}
\caption{Conical emission angle as a function of associated particle $\pt$ from STAR~\cite{3part}.}
\label{fig:3part_angle}
\end{minipage}
\end{figure}

Figure~\ref{fig:3part} shows the three-particle azimuthal correlation results in d+Au and central 12\% Au+Au collisions from STAR~\cite{3part}. The d+Au data show well-defined back-to-back di-jet correlation peaks. The central Au+Au data show a near-side peak and rich structures on the away-side. The back-to-back peak on the away-side is significantly weaker than in d+Au and appears elongated along and main diagonal axis. Additional peaks are observed on the off-diagonal axis. These peaks are evidence of conical emission. This is because particle emission on a cone around the away-side jet axis leads to pairs of correlated hadrons dominantly populating at the cone angle from $\pi$, with equal probability to be far apart (symmetric about $\pi$) and close together~\cite{3part}. This yields four equal magnitude peaks, two along the off-diagonal axis and two along the diagonal with equal distance from $\pi$. Deflected jets scenario adds contributions only along the diagonal axis (which results in stronger diagonal peaks than off-diagonal peaks). The off-diagonal peaks are the unique signature of conical emission. 

Conical emission can be generated by Mach-cone shock-waves or $\check{\rm C}$erenkov gluon radiation. Mach-cone angle depends only on the medium speed of sound and does not depend on the particle $\pt$; $\check{\rm C}$erenkov radiation angle depends on the medium index of refraction and thus generally depends on the particle $\pt$. Figure~\ref{fig:3part_angle} shows the conical emission angle fit to the three-particle azimuthal correlation data as a function of associated particle $\pt$. The emission angle is rather independent of $\pt$, favoring the physics scenario of Mach-cone shock-waves.

In the Mach-cone scenario, the emission angle has imprinted the speed of sound of the medium. However, the measured emission angle may be affected by medium expansion, as indicated by phenomenological studies such as those in Ref.~\cite{Renk}. 
Further theoretical and experimental work are needed in order to make connections between the measured conical emission angle and the medium speed of sound.

\section{The Ridge}

\begin{figure}[htb]
\begin{minipage}{0.4\textwidth}
\includegraphics[width=\textwidth]{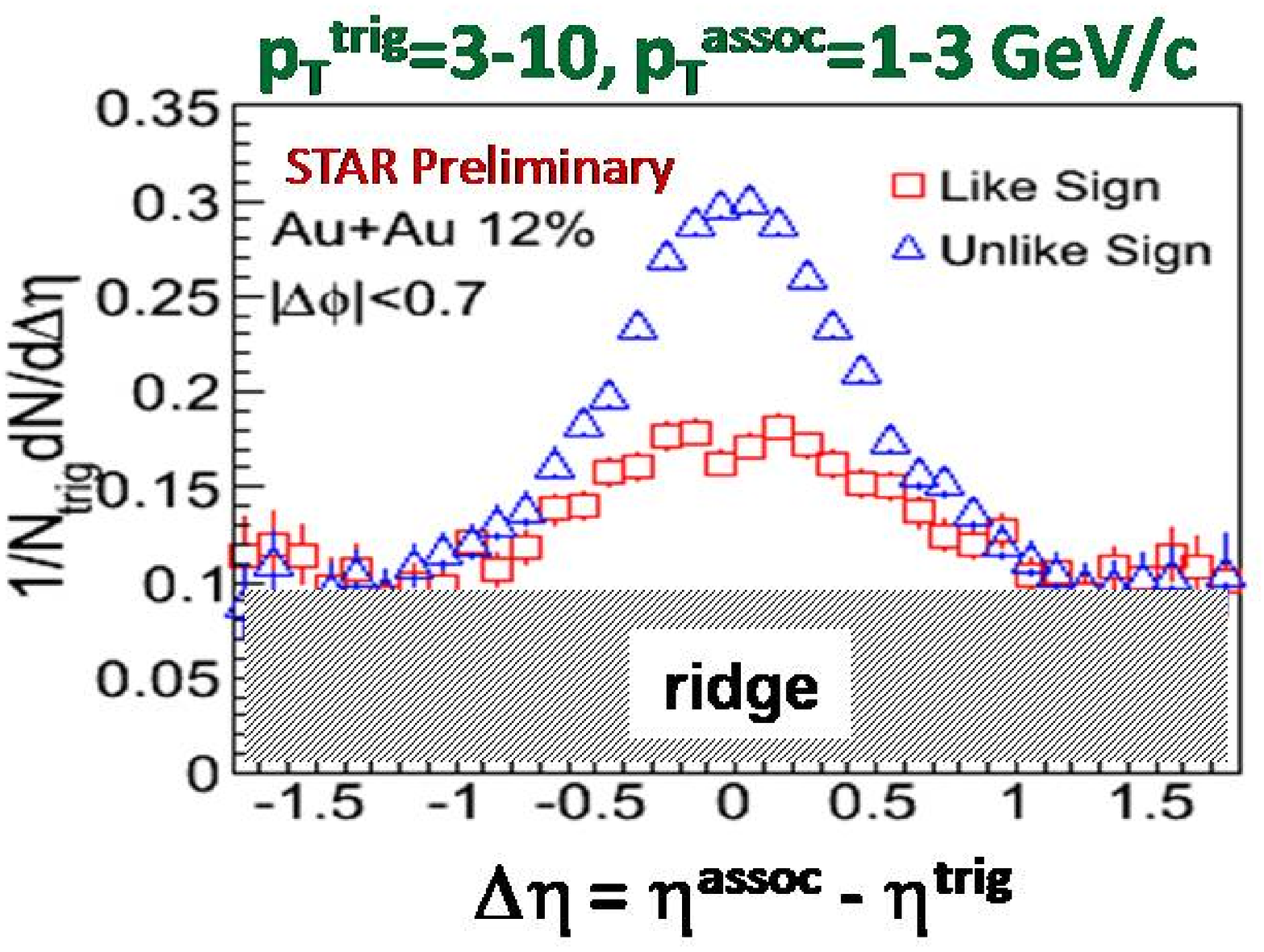}
\vspace*{-0.5in}
\caption{Background-subtracted near-side dihadron $\deta$ correlation from STAR~\cite{Pawan}. Background is normalized by {\sc zyam} at $\dphi\approx 1$.}
\label{fig:deta}
\end{minipage}
\hspace*{0.02\textwidth}
\begin{minipage}{0.58\textwidth}
\includegraphics[width=\textwidth]{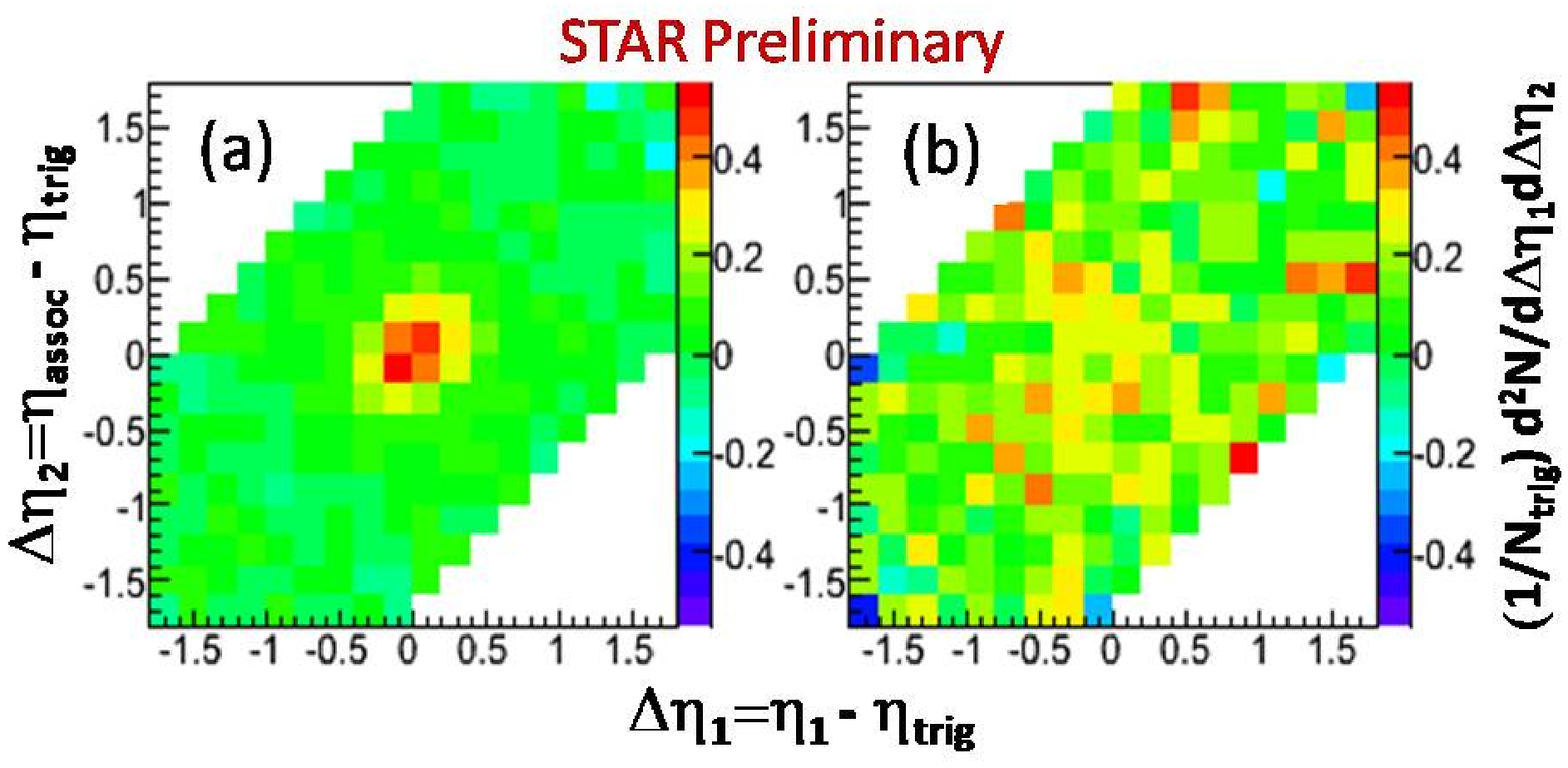}
\vspace*{-0.5in}
\caption{Three-particle pseudorapidity correlation in (a) d+Au and (b) central 12\% Au+Au collisions from STAR~\cite{Pawan}. Trigger and associated particle $\pt$ ranges are 3-10~\gev\ and 1-3~\gev, respectively. The correlations are for near-side $|\dphi|<0.7$, and corrected for 3-particle $\deta$ acceptance.}
\label{fig:3part_deta}
\end{minipage}
\end{figure}

Insights about jets may be gained by investigating the pseudorapidity $\deta$ correlation of associated particles within a narrow $\dphi$ region from the trigger particle. It was found that not all those correlated particles in narrow $\dphi$ (above the {\sc zyam} background) were also correlated with the trigger particle in narrow $\deta$. A significant fraction of those particles were found to be broadly distributed in $\deta$ in central Au+Au collisions~\cite{jetspec,Putschke} (see Fig.~\ref{fig:deta}). The observation of this so-called ridge has generated great interest. The properties of ridge particles, such as their particle compositions, are similar to those of inclusive particles~\cite{Nattrass}, however the origin of the ridge is presently not understood. Various theoretical models have been proposed, including longitudinal flow push~\cite{Armesto}, recombination between thermal and shower partons at intermediate $\pt$~\cite{Hwa2}, broadening of quenched jets in turbulent color fields~\cite{Majumder},  elastic collision between hard and medium partons (momentum kick)~\cite{Wong}, and QCD bremsstrahlung radiation or excess of particles due to Glasma flux tube fluctuation boosted by transverse flow~\cite{Voloshin,Shuryak,Dumitru,Dusling}.

Production of correlated particles in all these models can be broadly divided into three categories: (i) particles from jet fragmentation in vacuum which generate a jet-cone peak in dihadron correlation, (ii) particles from gluon radiation affected by the medium and diffused broadly in $\deta$ which generate the ridge, and (iii) particles from flux tube decays which can span many units of rapidity. However, the qualitative features of dihadron correlations are all similar from these models. In order to distinguish the models, STAR has carried out a three-particle pseudorapidity correlation analysis in $\deta$-$\deta$.

Figure~\ref{fig:3part_deta} shows the background-subtracted three-particle correlation in ($\deta_1$,$\deta_2$) space. A jet-fragmentation peak is clearly seen in d+Au data with no other contributions. In central Au+Au collisions, a peak at (0,0) is also observed but it is broadened and is atop a pedestal. The peak corresponds to the jet-like peak in dihadron correlation, and the pedestal is due to ridge correlation. However, the ridge and the jet-like component are not separated.

\begin{figure}[htb]
\centerline{\includegraphics[width=0.8\textwidth]{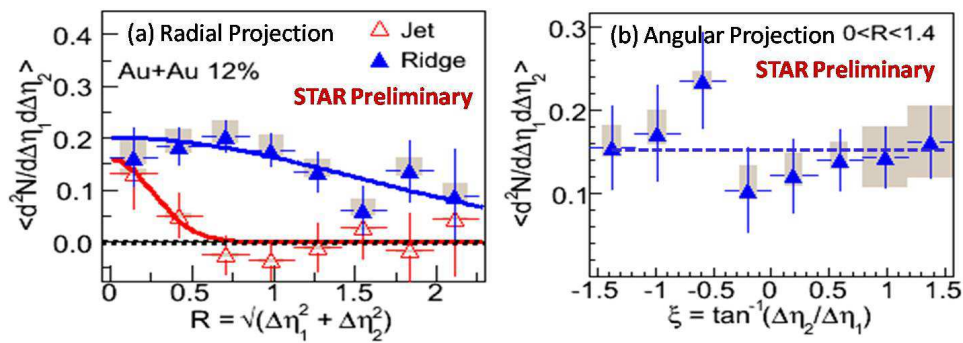}}
\vspace*{-0.4in}
\caption{Average three-particle pseudorapidity correlation signals of the ridge and the jet-like component in central 12\% Au+Au collisions as a function of (a) $R=\sqrt{\deta_1^2+\deta_2^2}$ and (b) $\xi=\tan^{-1}(\deta_2/\deta_1)$ within $0<R<1.4$. The curves in (a) are Gaussian fits.}
\label{fig:proj}
\end{figure}

Jet fragmentation has a charge ordering property. The probability to fragment into three same-sign hadrons at our energy is negligible. Any correlation in three same-sign hadron triplets is therefore dominated by the ridge correlation. This can be used to separate the ridge and the jet-like component. Figure 3(a) shows the average three-particle pseudorapidity correlation signals separately for the ridge and the jet-like component as a function of $R=\sqrt{\deta_1^2+\deta_2^2}$ in central 12\% Au+Au collisions. The ridge signal is consistent with constant. A Gaussian fit gives a width of $1.53\pm0.41$. On the other hand, the jet-like component is narrow with a Gaussian width of $0.25\pm0.09$. In order to investigate structures in the ridge, Fig.~\ref{fig:proj}(b) shows the average three-particle pseudorapidity correlation signal of the ridge within $0<R<1.4$ as a function of $\xi=\tan^{-1}\deta_2/\deta_1$. The data are constant with a uniform distribution, suggesting that the ridge particles are uncorrelated in $\deta$ not only with trigger particle but also between themselves. The ridge appears to be uniform event-by-event. In addition, no evidence is found for horizontal and vertical strips in the $\deta$-$\deta$ correlations which would correspond to associated pairs of a jet-like particle and a ridge particle. 

Most models proposed to explain the ridge in dihadron correlation fail to describe the event-by-event uniform ridge indicated by the three-particle pseudorapidity correlation results. The Glasma flux tube model~\cite{Dusling} predicts a structureless ridge in three-particle pseudorapidity correlation, consistent with data. However, the absence of correlation between the jet-like component and the ridge remains an open question.

\section{Summary}

Jet-like dihadron correlations with high-$\pt$ trigger particles in central heavy-ion collisions at RHIC are strongly modified from those in $pp$ collisions. The change in the away-side correlation structure is dramatic due to the maximal pathlength of the medium the away-side parton has to traverse. Three-particle azimuthal correlations in $\dphi$-$\dphi$ indicate that the away-side novel correlation structure is due to conical emission of correlated hadrons, consistent with Mach-cone shock-waves. Extraction of medium properties such as the speed of sound needs further theoretical and experimental work. 

A long range correlation in pseudorapidity, the ridge, is observed in central heavy-ion collisions. The particle compositions in the ridge are similar to the bulk medium. Three-particle pseudorapidity correlations in $\deta$-$\deta$ indicate that the ridge particles are uniform event-by-event, suggesting that the ridge and the jet-like component triggered by a high-$\pt$ particle may originate from different physics. The ridge may be the medium itself. Color flux tube fluctuation promises to be a viable explanation for the ridge. The apparent lack of correlation between the ridge and the jet-like component remains an open question.

\section*{Acknowledgments}
The author thanks the conference organizers for invitation and for the stimulating conference. This work is supported by U.S. Department of Energy under Grant DE-FG02-88ER40412.

\end{document}